\documentclass[10pt,aps,prd,twocolumn, nofootinbib,superscriptaddress]{revtex4-1}
 
\usepackage{amssymb,amsmath,accents,mathrsfs}
\usepackage{aas_macros} 
\usepackage{subcaption,verbatim}
\usepackage{graphicx}
\usepackage{color,units}
\usepackage{hyperref}
\usepackage{bm}
\usepackage{listings}
\usepackage{soul} 
\usepackage[normalem]{ulem}
\hypersetup{
    bookmarks=true,         
    unicode=false,          
    pdftoolbar=true,        
    pdfmenubar=true,        
    pdffitwindow=false,     
    pdfstartview={FitH},    
    pdfauthor={Boris Goncharov},     
    colorlinks=true,       
    linkcolor=blue,          
    citecolor=red,        
    urlcolor=blue,
}
\graphicspath{{}}

\begin{document}

\definecolor{dkgreen}{rgb}{0,0.6,0}
\definecolor{gray}{rgb}{0.5,0.5,0.5}
\definecolor{mauve}{rgb}{0.58,0,0.82}

\lstset{frame=tb,
  	language=Matlab,
  	aboveskip=3mm,
  	belowskip=3mm,
  	showstringspaces=false,
  	columns=flexible,
  	basicstyle={\small\ttfamily},
  	numbers=none,
  	numberstyle=\tiny\color{gray},
 	keywordstyle=\color{blue},
	commentstyle=\color{dkgreen},
  	stringstyle=\color{mauve},
  	breaklines=true,
  	breakatwhitespace=true
  	tabsize=3
}

\title{Inferring fundamental spacetime symmetries with gravitational-wave memory: from LISA to the Einstein Telescope}

\author{Boris Goncharov}%
 \email{boris.goncharov@me.com}
\affiliation{Gran Sasso Science Institute (GSSI), I-67100 L'Aquila, Italy}
\affiliation{INFN, Laboratori Nazionali del Gran Sasso, I-67100 Assergi, Italy}
\affiliation{Max Planck Institute for Gravitational Physics (Albert Einstein Institute), D-30167 Hannover, Germany}
\affiliation{Leibniz Universität Hannover, D-30167 Hannover, Germany}

\author{Laura Donnay}%
\affiliation{International School for Advanced Studies (SISSA), Via Bonomea 265, 34136 Trieste, Italy}
\affiliation{INFN, Sezione di Trieste,
Via Valerio 2, 34127 Trieste, Italy}

\author{Jan Harms}%
\affiliation{Gran Sasso Science Institute (GSSI), I-67100 L'Aquila, Italy}
\affiliation{INFN, Laboratori Nazionali del Gran Sasso, I-67100 Assergi, Italy}

\date{\today}

\begin{abstract}
We revisit gravitational wave (GW) memory as the key to measuring spacetime symmetries, extending beyond its traditional role in GW searches.
In particular, we show how these symmetries may be probed via displacement and spin memory observations, respectively. 
We further find that the Einstein Telescope's (ET) sensitivity enables constraining the strain amplitude of a displacement memory to 2\% and that of spin memory to 22\%. 
Finally, we point out that neglecting memory could lead to an overestimation of measurement uncertainties for parameters of binary black hole (BBH) mergers by about 10\% in ET.
\end{abstract}

\maketitle

\textit{Introduction.}---General Relativity (GR) predicts that, in addition to the oscillatory contributions to gravitational wave (GW) strain, there are also non-oscillatory contributions which are referred to as gravitational memory~\citep{Zel'dovichPolnarev1974,Christodoulou:1991cr,Blanchet:1992br,Thorne1992}. 
By Strominger and Zhiboedov~\citep{StromingerZhiboedov2014}, GW memory has been shown to represent one of the three corners of the so-called ``infrared triangle''. The latter establishes universal relations between the memory effect, Weinberg's soft graviton theorem in quantum field theory \citep{Weinberg:1965nx} and the Bondi-Metzner-Sachs (BMS) symmetry group of asymptotically flat spacetimes~\citep{Bondi:1962px,Sachs:1962wk}. The three seemingly unrelated subjects turn out to be three equivalent ways of characterizing physics at very long distances and the recent exploration of their connection has shed new light on the fascinating universal features of gravity~\citep{Strominger:2017zoo}.

The \textit{standard} BMS group describes the asymptotic symmetries of flat spacetime at null infinity, \textit{i.e.} as seen by observers infinitely far away from the gravitational field of interest.
The authors of~\citep{Bondi:1962px,Sachs:1962wk} initially expected that symmetries of such spacetimes would represent the Poincar{\'e} group, with ten conserved quantities -- referred to as charges -- being the total energy, the momentum, the angular momentum, and the position of the center of mass.
However, they ended up discovering a superset of the Poincar{\'e} group which also contains an infinite amount of so-called supermomentum charges.
These charges are associated with the symmetries known as \emph{supertranslations}, as they extend the usual group of Poincar{\'e} translations.

BMS charges are conserved in the absence of fluxes of radiation to null infinity.
In the presence of radiation, they rather obey flux-balance laws (among which one finds the Bondi mass loss formula \citep{Bondi:1962px}). The displacement memory arises as the permanent shift of the asymptotic shear after the passage of GWs. This shift can be equivalently described as a transition between two different asymptotic BMS frames related by a supertranslation \citep{Ashtekar:2014zsa,StromingerZhiboedov2014}. 

Later studies proposed more relaxed fall-off conditions for the metric near null infinity and have been shown  to lead to new GW memory terms.
The symmetry group which we refer to as the \textit{extended} BMS group includes additional ``superrotation'' symmetries that correspond to supermomentum and superspin charges~\citep{Barnich:2009se,Barnich:2010eb,Kapec:2014opa}.
They can be related to a new GW memory known as the spin memory~\citep{Pasterski:2015tva,Nichols:2018qac,Himwich:2019qmj,Flanagan:2015pxa}, which manifests as a relative time delay between light rays in counter-orbiting trajectories.
Additional symmetry groups were proposed later~\citep{Campiglia:2015yka,Freidel:2021fxf,Godazgar:2018qpq}, associated additional memory-type effects were discussed in~\citep{Compere:2018ylh,Seraj:2021rxd,Godazgar:2022pbx,Tahura,Hou:2020tnd}.  
Given the variety of proposals for the largest nontrivial asymptotic symmetry group, it is of great
interest to determine which set of BMS symmetries can be accounted for by observational data.

Previous experimental studies mainly concerned the detection of memory.
Current ground-based laser interferometers, the Advanced LIGO\footnote{Laser Interferometer Gravitational-wave Observatory (LIGO).}~\citep{LIGOScientificCollaborationAasi2015} and the Advanced Virgo~\citep{AcerneseAgathos2015}, may be able to detect the displacement component of the memory, as suggested by Lasky \textit{et al}.~\citep{LaskyThrane2016}.
Searches for displacement memory in multiple LIGO-Virgo signals has been carried out by~\citep{HubnerTalbot2020,HubnerLasky2021,CheungLasky2024}.
The authors found that the detection requires between a few hundred and a few thousand compact binaries detected by the LIGO and Virgo at design sensitivity.
Grant and Nichols~\citep{GrantNichols2023} developed a framework to forecast the detection of displacement and spin memory, evaluating the prospects of resolving spin memory with the Cosmic Explorer~\citep[CE, ][]{ReitzeAdhikari2019}.
With Pulsar Timing Arrays, the displacement memory~\citep{WangHobbs2015,ArzoumanianBrazier2015} may be detectable at $2\sigma$~\citep{vanHaasterenLevin2010}.
Prospects of memory detection are brighter~\citep{Favata2010,SunShi2023} with the space-based GW detector LISA\footnote{Laser Interferometer Space Observatory (LISA).}~\citep{Amaro-SeoaneAudley2017}.
There are additional potentially observable effects that have previously been referred to as ``memory'', e.g. ``velocity memory''~\citep{GrishchukPolnarev1989}.
However, they do not correspond to asymptotic symmetries, and so the term ``persistent observables'' was coined by Flanagan \textit{et al.} to describe this more general class of phenomena~\citep{FlanaganGrant2019}.
It is also worth mentioning that several publications discussed the merits of displacement memory in breaking GW parameter estimation degeneracies~\citep{GasparottoVicente2023,XuRossello-Sastre2024}, identifying low-mass BBHs~\citep{EbersoldTiwari2020}, and distinguishing black holes and neutron stars~\citep{YangMartynov2018,TiwariEbersold2021,LopezTiwari2024}.

In this Letter, we evaluate the prospects of characterizing memory and performing model selection to determine which set of symmetries describes the spacetime we live in.
We simulate GW signals from BBH mergers in different scenarios of spacetime at null infinity. 
We then evaluate the odds of distinguishing between the aforementioned scenarios with ground-based and space-based GW detectors.

\textit{Models.}---The boundary of flat spacetimes (null infinity $\mathscr I$) is a lightlike hypersurface parametrized by a null time $u=t-r$ and a two-dimensional sphere $S^2$, called the celestial sphere. The latter has a metric given by $\gamma_{AB}dx^A dx^B=d\theta^2+\sin^2\theta d\phi^2$ in terms of the usual spherical coordinates $x^A=(\theta,\phi)$. BMS transformations are generated by vector fields $\xi$ of the following form\footnote{There are also subleading terms in $1/r$; we only wrote here the restriction at $\mathscr I$ of the BMS vector field.}:
\begin{equation}
\xi=\left[T(x^A)+\frac{u}{2}D_A Y^A(x^B)\right]\partial_u+Y^A(x^B)\partial_A\,,
\end{equation}
where $D_A$ denotes the covariant derivative associated to $\gamma_{AB}$. Our models correspond to different choices of the supertranslation and superrotation generators $T$, and $Y^A$, respectively. 
In both Poincar{\'e} and standard BMS groups, the vector fields $Y^A$ are $\ell=1$ vector spherical harmonics generating the usual 6 Lorentz transformations. The infinite-dimensional nature  of the standard BMS group stems from the fact that the function $T(x^A)$ is allowed to be an arbitrary (smooth) function of the sphere angles (in contrast, in the Poincar{\'e} group, $T$ is restricted to be a linear combination of $\ell=0, 1$ spherical harmonics). 

The extended BMS group adds superrotations \citep{Barnich:2009se}, where $Y^A$ are taken to be \emph{local} conformal Killing vectors (CKV) of the sphere.  The key is to allow for singularities at isolated points on the celestial sphere\footnote{This implies that supertranslations can now also be singular.}. While this might seem odd from a conservative GR point of view (see, however, \citep{Strominger:2016wns} for a physical interpretation of superrotation symmetries), there is a strong motivation coming from a field theorist's perspective. Indeed, the algebra spanned by local CKV is nothing but the Virasoro algebra, an infinite-dimensional algebra ubiquitous in string theory and two-dimensional conformal field theory (CFT). The realization that Virasoro symmetries could arise as part of the symmetry group acting on the celestial sphere kicked off the ambitious Celestial Holography program~\citep{Pasterski:2021raf} aiming to provide a holographic description of quantum gravity in flat spacetime in terms of a CFT on the celestial sphere. 

Finally, let us mention that there exists a fourth option which is the generalized BMS group~\citep{Campiglia:2015yka}. It is characterized by the fact that $Y^A$ can be chosen to be any smooth vector fields (\textit{i.e.}, they generate the full diffeomorphisms of the sphere, Diff$(S^2)$). These generalized symmetries can now deform the boundary sphere metric and require enlarging the gravitational phase space; see \citep{Compere:2018ylh} for proposed associated memory effects. The four models are summarized in Table~\ref{tab:models}. 

\begin{table*}[!htb]
\caption{\label{tab:models}Models. We simulate and test the presence of memory terms based on the BMS balance laws~\citep{MitmanIozzo2021}. Based on the (non-)observation of a set of memory effects, we then identify a correct symmetry group.}
\begin{ruledtabular}
\begin{tabular}{l c c c c }
~ & Poincar{\'e} & Standard BMS & Extended BMS & Generalized BMS \\ \hline
Asymptotic symmetries & 4 translations & supertranslations & supertranslations & supertranslations \\ 
 & 6 Lorentz & 6 Lorentz & local CKV\footnote{local CKV field, as opposed to  global CKV (i.e. Lorentz transformations).} & Diff($S^2$)\\ 

Boundary metric  & $\gamma_{AB}$ & $\gamma_{AB}$ & $\gamma_{AB}$ locally\footnote{\textit{i.e} except at isolated points.} & not fixed \\ 
Memory terms & None & Displacement & Displacement, spin & Refraction, velocity kick\footnote{These terms are known to arise in the presence of impulsive gravitational waves~\citep{BhattacharjeeKumar2019}.} \\ 
\end{tabular}
\end{ruledtabular}

\end{table*}

We focus on the first three models of spacetime symmetries: the Poincar{\'e} spacetime, the \textit{standard} BMS, and the \textit{extended} BMS.
The Poincar{\'e} model indicates an absence of GW memory.
Therefore, it aligns with unlikely scenarios, \textit{e.g.}, ignoring GWs' cumulative contribution to Einstein's stress-energy tensor.
Detecting this scenario could stem from data analysis nuances, like model misspecification, so we still consider this model for generality.
The \textit{standard} BMS model entails observing only displacement memory. 
The relationship between superrotations and the spin memory is more subtle. 
Unlike displacement memory, which represents a shift to a supertranslated time frame, spin memory does not directly correspond to a spacetime ``superrotated'' from an early frame \citep{Nichols:2017rqr,Compere:2018ylh}.
Keeping this in mind, we will nevertheless consider the \textit{extended} BMS model as encompassing both displacement and spin memory observations.
We model GW strain over time $h(t)$ using the approximant from~\citep{VarmaField2019}.

\textit{Observations.}---Determining which spacetime symmetry group represents the universe we live in\footnote{While we live in a de Sitter universe, the approximation of an asymptotically flat universe is valid up to astrophysical scales smaller than cosmological scales~\citep{KehagiasRiotto2016}.} reduces to determining which set of memory terms is the best description of GW data.
In this Letter, we simulate measurements with:
\begin{itemize}
    \item Next-generation detectors, the Einstein Telescope (ET)~\citep{PunturoAbernathy2010}, and the Cosmic Explorer (CE)~\citep{ReitzeAdhikari2019};
    \item LISA, a proposed space-based instrument~\citep{ColpiDanzmann2024}. 
\end{itemize}

GW data analysis is based on the likelihood $\mathcal{L}(d|\theta)$ of the data $d$ as a function of model parameters $\theta$.
The data is fit to the model in the frequency ($f$) domain, $\tilde{h}(f)$.
The measurement of parameters is given by the posterior probability $\mathcal{P}(\theta|d)=\mathcal{L}(d|\theta)\pi(\theta)/\mathcal{Z}$.
The function $\pi(\theta)$ is a prior and $\mathcal{Z}=\int \mathcal{L}(d|\theta)\pi(\theta) d\theta$ is referred to as the evidence.
The Bayes factor $\mathcal{B}$ is the ratio of $\mathcal{Z}$ of the two models. 
We consider the prior odds of the models to be equal, thus $\mathcal{B}$ is also the Bayesian odds ratio. 
The priors are usually chosen such that they represent the observed distribution of $\theta$.
Because it has a negligible effect on our results, we set $\pi(\theta) = 1$, such that the posterior is driven entirely by the likelihood. 
We also employ the Fisher matrix approximation of the likelihood~\citep{DupletsaHarms2023}.
We evaluate $\mathcal{P}(\theta|d)$ for a full set of parameters that describe a BBH merger: binary component masses $m_{1,2}$, angular momenta $a_{1,2}$, GW phase $\phi$ and polarization $\psi$, orbital inclination to the line of sight $\theta_{jn}$, luminosity distance $D_\text{L}$, sky position and merger time.
We also introduce additional parameters, multiplicative factors of the predicted GW memory strain amplitude, $A_\text{d}$ for displacement memory and $A_\text{s}$ for spin memory, for the following reason.
Model selection requires evaluating the Bayesian evidence for both the correct model and the incorrect model.
Fitting an incorrect model to the data introduces systematic errors and, on average, a decrease in the maximum likelihood.
Introducing $A_{\text{d},\text{s}}$ allows measuring posterior probability, as well as the Bayesian evidence, for both the model where memory is present in the data ($A_{\text{d},\text{s}}=1$) and for the model where it is not present ($A_{\text{d},\text{s}}=0$), by considering slices along $A_{\text{d},\text{s}}$ of the full posterior.
Thus, $A_{\text{d},\text{s}}$ manifests as a Bayesian hyper-parameter. 
The implications of our approximations are discussed in the \hyperlink{supplementary_material}{Supplementary Material}.

\textit{LISA.}---Space-based interferometer LISA is to explore the GW universe between $10^{-4}$ and $1$ Hz in frequency. 
The loudest expected astrophysical sources for LISA are mergers of massive BBHs, which may be detected with a rate of a few per year~\citep{BarausseDvorkin2020}.
Frequency-domain strain $\tilde{h}(f)$ is projected onto the three LISA data channels~\citep{TintoEstabrook2002}, using detector specifications from~\citep{Amaro-SeoaneAudley2017}.
We simulate frequency series from $f_\text{min}=2^{-13}~\text{Hz} \approx 1.2\times10^{-4}~\text{Hz}$ to $f_\text{max}=2^{-6}~\text{Hz} \approx 1.6 \times 10^{-2}~\text{Hz}$, with a spacing $\Delta f=f_\text{min}$.

\begin{figure}[!htb]
    \centering
    \includegraphics[width=\columnwidth]{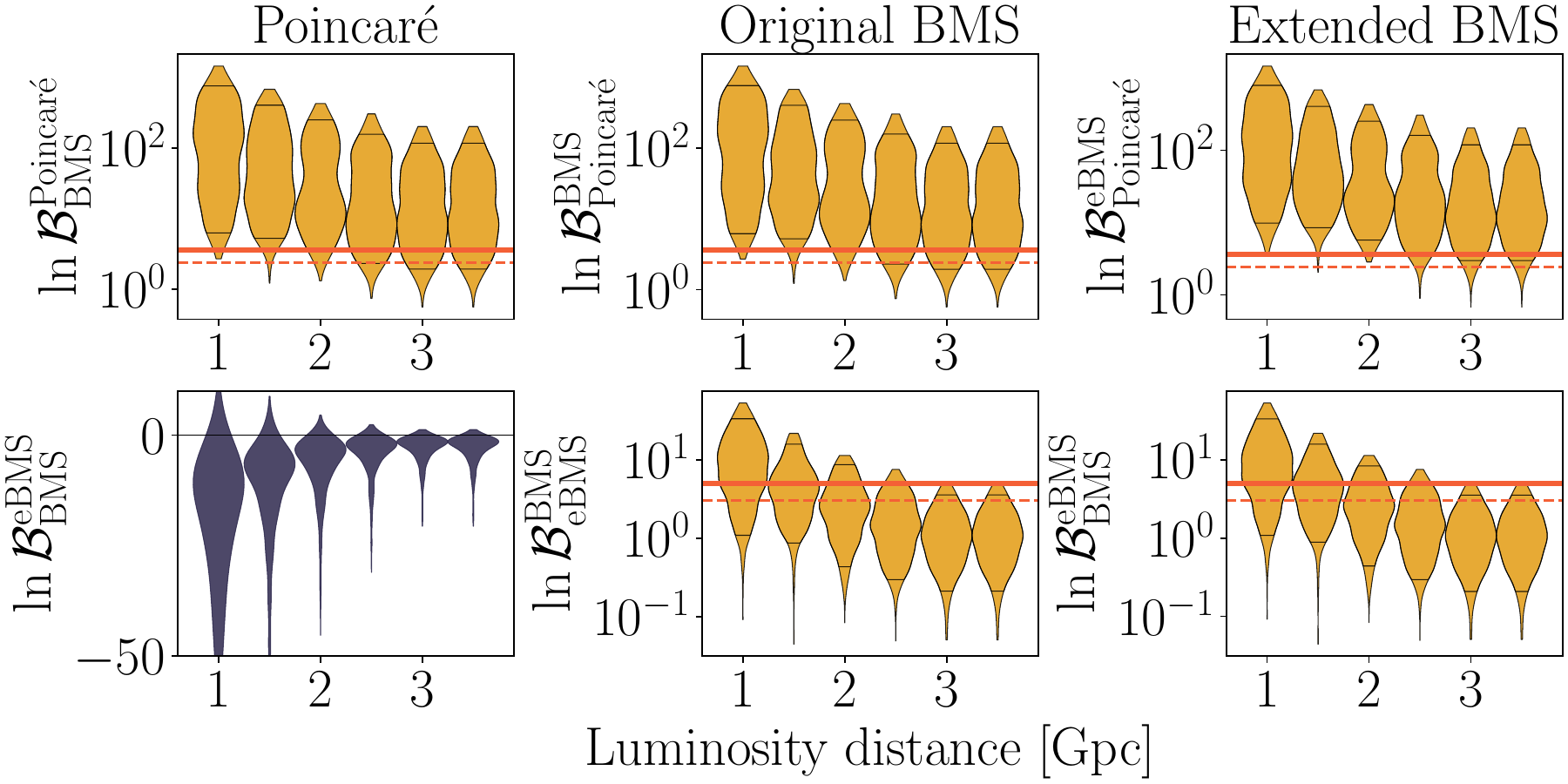}
    \caption{Model selection with LISA based on single non-spinning BBHs. Panels show the inferred log odds between symmetry groups, $\ln\mathcal{B}$, marginalized over realizations of GW parameters other than component masses ($10^6~M_\odot$).  Panel titles denote correct symmetry groups in simulated scenarios. Horizontal dashed lines correspond to $\ln\mathcal{B}=3$ (strong evidence), solid lines correspond to $\ln\mathcal{B}=5$ (detection). 
    }
    \label{fig:lisa}
\end{figure}

We examine a non-spinning BBH with equal component masses of $10^6~M_\odot$, positioned at a $D_\text{L}$ between $1$ and $3$ Gpc.
LISA is expected to detect such a signal during the mission lifetime of 4-6 years.
We simulate stochastic realizations of this observation in universes described by different asymptotic spacetime symmetry groups, marginalizing over sky position, merger time, $\phi$, $\psi$, and $\theta_{jn}$.
In Figure~\ref{fig:lisa}, Bayesian odds between symmetry groups are shown against $D_\text{L}$.
In panels with log-y-axes, $5$-$95$ \% cumulative density levels are shown as horizontal black lines across the distributions.
The top three panels suggest that LISA could confirm or refute that spacetime at null infinity is described by the Poincar{\'e} group.
This result hinges solely on whether displacement memory is detected or not.
The bottom three panels correspond to choosing between original and extended BMS models, determined by LISA's capacity to detect spin memory. 
In the Poincar{\'e} scenario, selecting higher-order symmetries implies model misspecification (bottom left panel).
Comparing a large sample of real data with simulations could reveal such a case.
Finally, the two bottom right panels indicate that single-event-based inference of higher-order BMS symmetries is likely only for $10^6~M_\odot$ binaries within the $2$ Gpc range.

\textit{Ground-based detectors.}---Considering 1 year of observation of BBHs with (a) ET, (b) a network of ET and CE, we simulate frequency series from $f_\text{min}=10$ Hz to $f_\text{max}=1024$ Hz, with a spacing $\Delta f = 2^{-7}$ Hz.
Because single-event memory inference is less effective than with LISA, we combine data from multiple binary coalescences.
Because the observations are independent, the total log Bayes factor $\ln\mathcal{B} = \sum_i \ln\mathcal{B}_i$, where $i$ indexes each BBH observation~\citep{LaskyThrane2016,HubnerTalbot2020,HubnerLasky2021}. 
Note, summing log Bayes factors is similar to the widely-used procedure of adding SNRs in quadrature because $\ln\mathcal{L} \propto \text{SNR}^2$ (optimal SNR, frequentist statistic). 
Accordingly, $\ln\mathcal{B}$ should increase linearly with observation time, while SNR grows as the square root of time.
To simulate a realistic Bayesian model selection with $\phi$-$\psi$ degeneracy limiting our ability to determine the sign of the memory, we assign $\ln\mathcal{B}$ to zero for signals with optimal SNR of the $m$-odd part of the waveform $<2$~\citep{LaskyThrane2016,GrantNichols2023}.
We simulate two populations of BBHs up to $z=30$ based on the analysis of the second LIGO-Virgo transient catalog~\citep{AbbottAbbott2021b}.
Our main population is based on the median-posterior local merger rate of $23.9~\text{Gpc}^{-3}\text{yr}^{-1}$ and the pessimistic one is based on the lower limit at 95\% credibility, $15.3~\text{Gpc}^{-3}\text{yr}^{-1}$.
The mass distribution is described by the best-fit model and best-fit parameters from~\citep{AbbottAbbott2021b}.
Inference of spacetime symmetries does not suffer from astrophysical selection effects, so we focus on the best 1000 events that maximize the optimal SNR for the most elusive spin memory component.

\begin{figure}[!htb]
    \centering
    \includegraphics[width=\columnwidth]{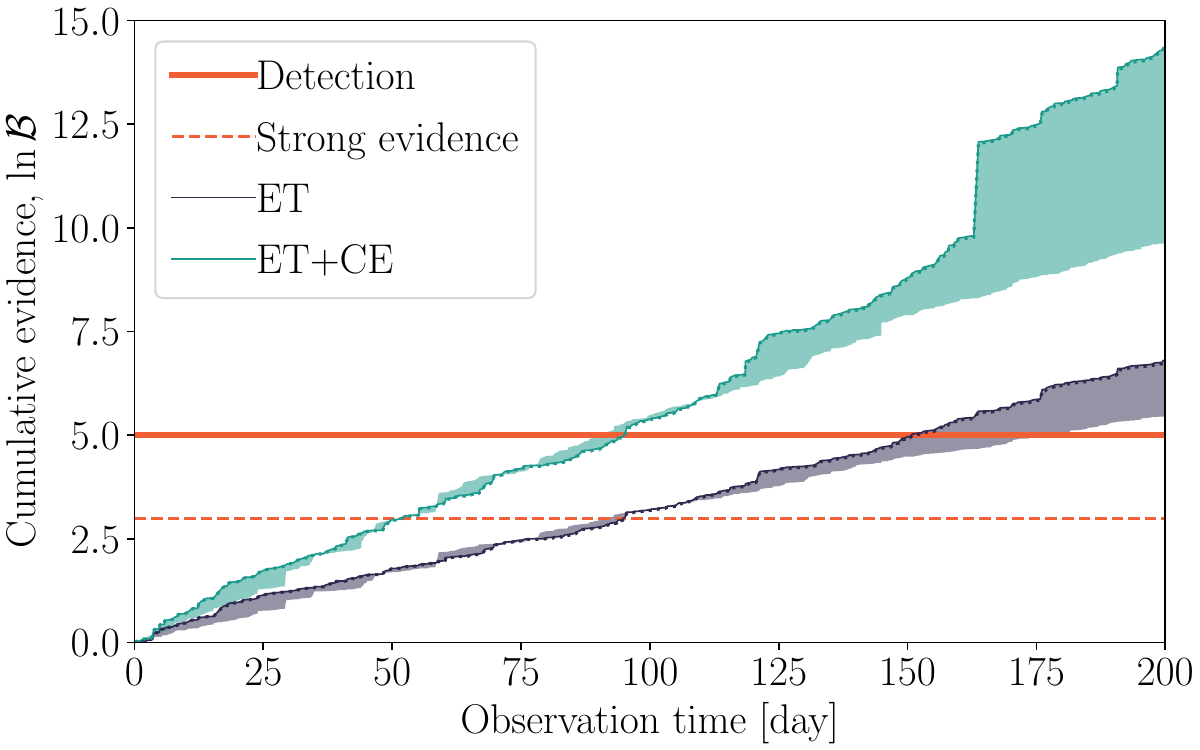}
    \caption{Model selection between the original BMS (dotted lines) and the extended BMS (solid lines) with ET and CE. Evidence for the simulated symmetry group $\ln\mathcal{B}$ is shown against the observation time.
    } 
    \label{fig:combine}
\end{figure}

Figure~\ref{fig:combine} shows how enhanced symmetries of the extended BMS may be inferred or ruled out with the ET and the joint ET-CE observations.
The outcome is due to the sensitivity of detector networks to spin memory.
For ET, it is sufficient to analyze around $100$ optimal events to make such a conclusion or to operate for $150$-$175$ days.
Whereas for an ET-CE network, this reduces to $10$-$30$ optimal events and $90$ days of observation, respectively.
Note, the presented Bayes factors are noise-averaged. 
The effect of Gaussian noise will introduce a range of the predicted values around our estimates.
Precisely, the linear growth of $\ln\mathcal{B}$ with the observation time will turn out to be a stochastic random-walk-like process~\citep{HubnerTalbot2020}.
We further support~\citep{LaskyThrane2016,HubnerTalbot2020} by finding that a network of LIGO-Virgo may be able to distinguish between the original BMS and the extended BMS in around a year, thanks to a sensitivity to displacement memory.
The result is dominated by $\approx 100$ optimal detections, further $1000$ detections do not add to the measurement because of the $\phi$-$\psi$ degeneracy. 

\textit{Constraining spacetime symmetries.}---Introducing $A_{\text{d},\text{s}}$ leads to their posterior distributions.
Previously, we used these posteriors solely for Bayes factor evaluation.
However, we propose examining $A_{\text{d},\text{s}}$ measurements themselves.
Any deviations from the values of $0$ and $1$ could be indications of model misspecification, indicating either new physics or unknown systematic errors~\citep{ScargleCao2021}.
Similarly to taking the product of $\mathcal{B}$, the posterior distribution given all available data $\mathcal{P}(A_{\text{d},\text{s}}|d)$ factorizes as a product of posterior distributions from individual GW events, marginalized over nuisance parameters.
Additionally, here, we simulate offsets in $A_{\text{d},\text{s}}$ due to Gaussian noise following Equation~5 in~\citep{SharmaHarms2020}.

\begin{figure}[!htb]
    \centering
    \includegraphics[width=\columnwidth]{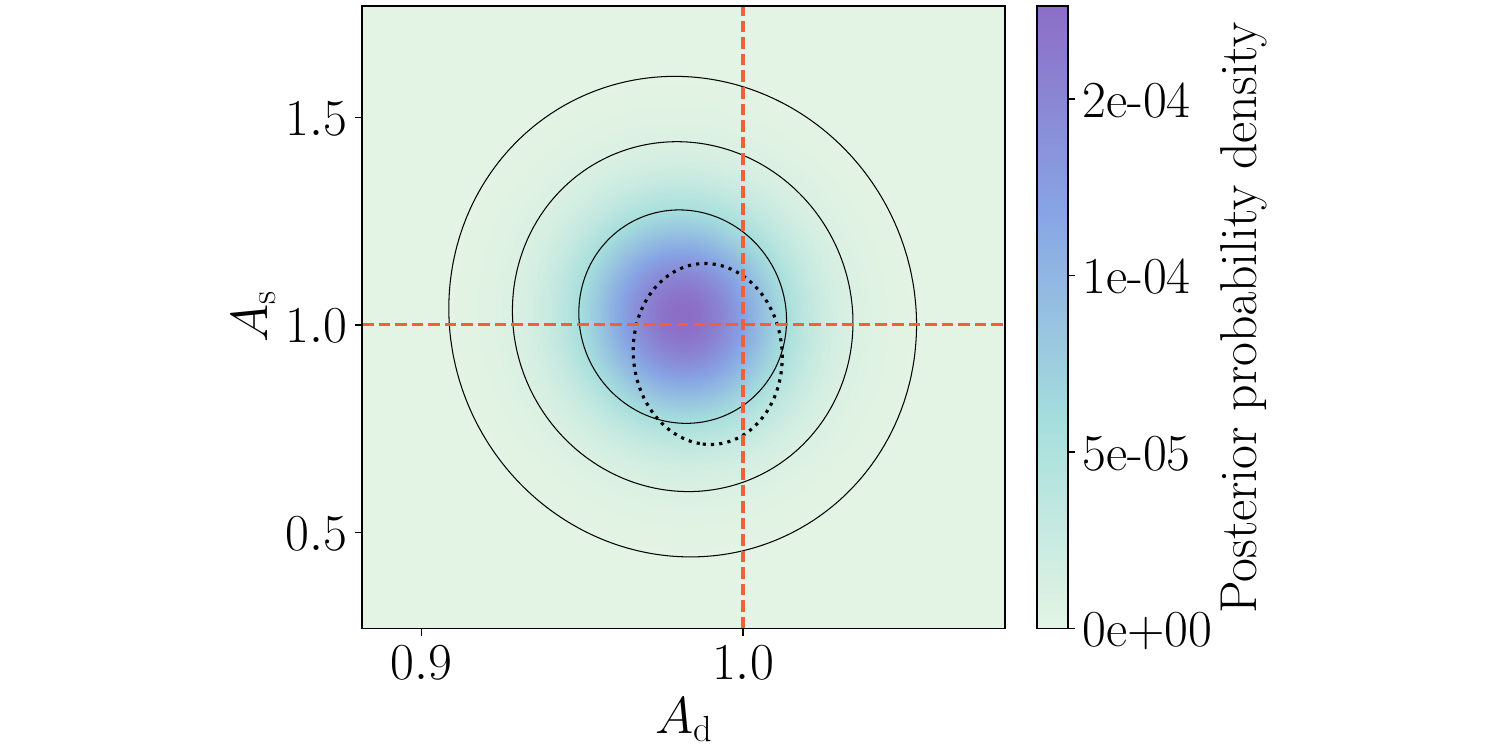}
    \caption{Relative memory amplitudes, $A_\text{d,s}$, with $10^3$ optimal BBHs observed by ET in a year, in the extended BMS scenario. Solid lines show $1,2,3$-$\sigma$ credible levels. Dotted lines show $1$-$\sigma$ credible levels when we include all BBH mergers up to a redshift $z \approx 1$ ($10^4$ events).
    } 
    \label{fig:pe_xg}
\end{figure}

Figure~\ref{fig:pe_xg} presents combined constraints on $A_{\text{d},\text{s}}$ for Poincar{\'e}, the original BMS, and the extended BMS models, using 1000 optimal BBHs observed by ET in a year.
The 1-$\sigma$ uncertainty of the displacement memory amplitude $A_{\text{d}}$ is $0.03$, whereas the respective uncertainty of the spin memory amplitude $A_{\text{s}}$ is $0.26$.
Additionally, combining data from all BBH mergers at redshifts $z \lessapprox 1$ with ET, totaling approximately $10^4$ events, further tightens $A_\text{d}$ uncertainty to $0.02$ and $A_\text{s}$ to $0.22$.
The best-fit $A_{\text{d,s}}$ are consistent with the simulated values. 
For comparison, the authors of a recent preprint~\citep{CheungLasky2024} have found $A_\text{d}<15$ using the data from the third LIGO-Virgo transient catalog.

Following ~\citep{GasparottoVicente2023,XuRossello-Sastre2024}, we explore whether GW memory terms enhance parameter estimation for BBH inspiral and merger in ET.
We find that while elusive spin memory has negligible impact, the inclusion of displacement memory in simulated scenarios significantly improves measurement accuracy. 
Below, we report the reduction of measurement uncertainties, on average, for the $4000$ optimally chosen BBH mergers from a year of observations: 
\begin{itemize}
    \item $\phi$: by $81\%$,
    \item $\psi$: by $74\%$,
    \item sky position: by $41$-$42$\%,
    \item merger time: by $41\%$,
    \item $D_\text{L}$: by $37\%$,
    \item $\theta_{jn}$ (degenerate with $D_\text{L}$): only by $4\%$,
    \item $m_{1,2}$: by $32$-$34\%$,
    \item $a_{1,2}$: by $12\%$.
\end{itemize}
For several loudest BBH mergers, systematic errors from assuming incorrect $A_\text{d,s}$ reach $1$-$2~\sigma$.
Our findings suggest that displacement memory -- which is expected to exist in nature -- would allow us to extract more information about the astrophysics of compact binary mergers.

\textit{Conclusion.}---
We simulated a year of BBH observations with ET and CE, along with single-event simulations using LISA, for three hypothetical spacetime symmetry scenarios.
Determining the symmetry group that characterizes our universe depends on identifying the presence or absence of displacement and spin memory.
Averaging over Gaussian noise, we discover that ET could select between the standard BMS group and the extended BMS group in less than a year. 
Simultaneous GW observations with ET and CE further reduce this timescale and increase the likelihood of observing fairly strong ($\ln\mathcal{B} \approx 2$) evidence for spin memory in individual signals.
Additionally, LISA will almost certainly distinguish between the standard BMS spacetime and the Poincar{\'e} spacetime when observing a $10^6~M_\odot$ binary at $<3.5~\text{Gpc}$.
LISA also has a $\approx 50\%$ chance of distinguishing between the standard BMS and the extended BMS scenario if such a binary is observed at $<1.5~\text{Gpc}$.
Moreover, we test the idea of measuring the amplitude of displacement memory, $A_\text{d}$, and of the spin memory, $A_\text{s}$.
This may be employed as a test of spacetime symmetries, akin to tests of General Relativity~\citep{HeisenbergYunes2023}.
This could assist in identifying data analysis problems and model misspecification, \textit{i.e.} situations where the actual asymptotic spacetime symmetry group differs from our expected groups.
For example, a deviation from the expected $A_\text{d,s}$ may occur for BBH mergers at high redshifts~\citep{NgGoncharov2023}, where the approximation of asymptotically flat spacetime is no longer valid~\citep{KehagiasRiotto2016}. 
Thus, measuring $A_\text{d,s}$ will prove useful in memory searches by LIGO-Virgo~\citep{HubnerTalbot2020,HubnerLasky2021,CheungLasky2024}, as well as in future experiments.
We assess constraints on $A_\text{d,s}$ against the extended BMS predictions scaled to 1.
ET with $\approx 10^4$ optimal BBHs ($\approx 5\%$ of BBHs observed in a year) would constrain $A_\text{d}$ to $0.02$ and $A_\text{s}$ to $0.22$ at $1\sigma$.
Finally, displacement memory improves parameter estimation in ET and CE's loudest signals, while its omission could cause systematic $1$-$2\sigma$ errors.
This underscores the need for gravitational waveform models with memory~\citep{YooMitman2023} for ET, CE, and LISA.

We thank Paul Lasky, Eric Thrane, Geoffrey Comp\`ere, Keefe Mitman, and Leo Stein for useful discussions. 
We also appreciate early feedback on the manuscript from Tjonnie Li and Nathan Johnson-McDaniel.
We thank the organizers and speakers of the GW memory workshop at the Queen Mary University of London for an opportunity to touch base with the cutting-edge GW memory research before the completion of this work. 
Finally, we thank OzGrav, The Australian Research Council Centre of Excellence for Gravitational Wave Discovery, for providing the opportunity to utilize the OzSTAR Australian national facility (high-performance computing) at Swinburne University of Technology. 
BG is supported by the Italian Ministry of Education, University and Research within the PRIN 2017 Research Program Framework, n. 2017SYRTCN.
L.D. is supported by the European Research Council (ERC) Project 101076737 -- CeleBH. Views and opinions expressed are however those of the author only and do not necessarily reflect those of the European Union or the European Research Council. Neither the European Union nor the granting authority can be held responsible for them.
L.D. is also partially supported by INFN Iniziativa Specifica ST\&FI.
See \hyperlink{supplementary_material}{Supplementary Material} with Refs.~\citep{Braginsky:1986ia,Thorne:1986iy,PayneHourihane2022,TheLIGOScientificCollaborationtheVirgoCollaboration2021,HallKuns2021,AbbottAbbott2021a,MitmanMoxon2020,AbbottAbbott2019,IsloSimon2019,KagraCollaborationAkutsu2019,FlanaganNichols2017,PollneyReisswig2011,Vallisneri2008,BlanchetDamour1992,BraginskyThorne1987} for an extended set of figures and a more detailed description of the memory modeling, data analysis, and results.

\bibliography{mybib}{}
\bibliographystyle{unsrtnat}


\newpage

\onecolumngrid

\begin{center}
\textbf{\normalsize \hypertarget{supplementary_material}{SUPPLEMENTARY MATERIAL}}
\end{center}

\vspace{2\baselineskip} 

\twocolumngrid

\section{\label{sec:figures} Extended figures}

\begin{figure*}[!htb]
    \centering
    \includegraphics[width=\textwidth]{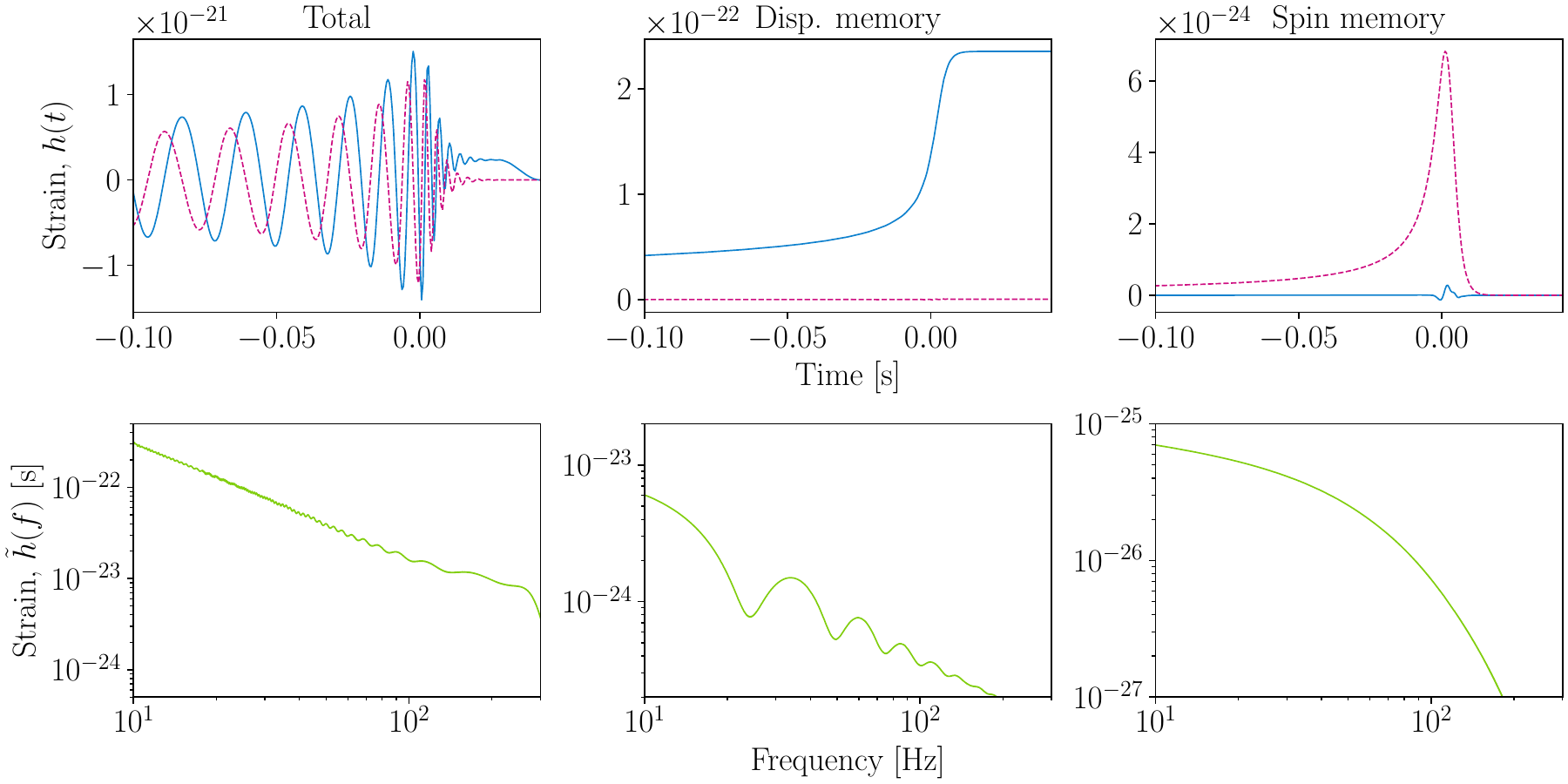}
    \vspace{-1.2\baselineskip} 
    \caption{Demonstration of the GW memory contribution to strain from a merger of two non-spinning BBHs in the extended BMS scenario, $(m_1,m_2,\theta_{jn},z)=(30~M_\odot,30~M_\odot,\pi/3,0.06)$. Solid lines show $h_+$, dashed lines show $h_\times$. The effect of windowing on $h(t)$ is shown only in the left top panel as a decline in strain after the merger at $t=0$~s.
    }
    \label{fig:mem_wave_sup}
\end{figure*}

\begin{figure*}[!htb]
    \centering
    \begin{subfigure}[b]{0.49\textwidth}
        \includegraphics[width=\textwidth]{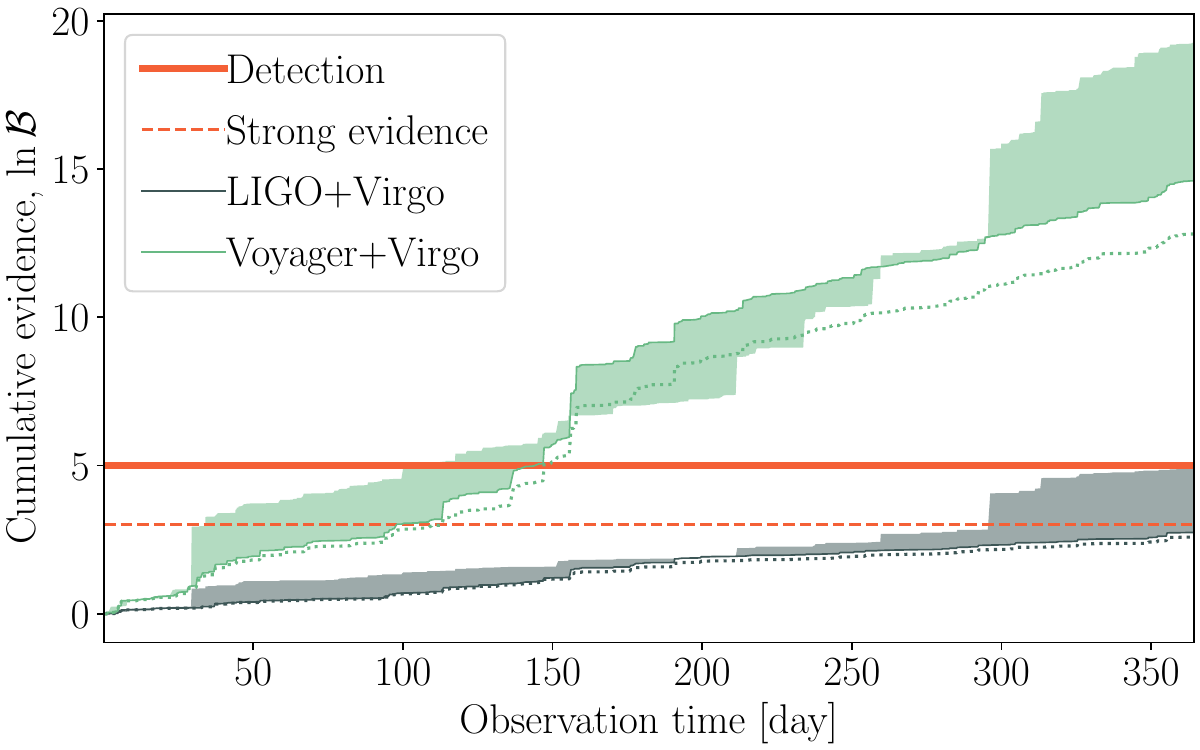}
        \label{fig:combine:lvktime}
    \end{subfigure}
    \begin{subfigure}[b]{0.49\textwidth}
        \includegraphics[width=\textwidth]{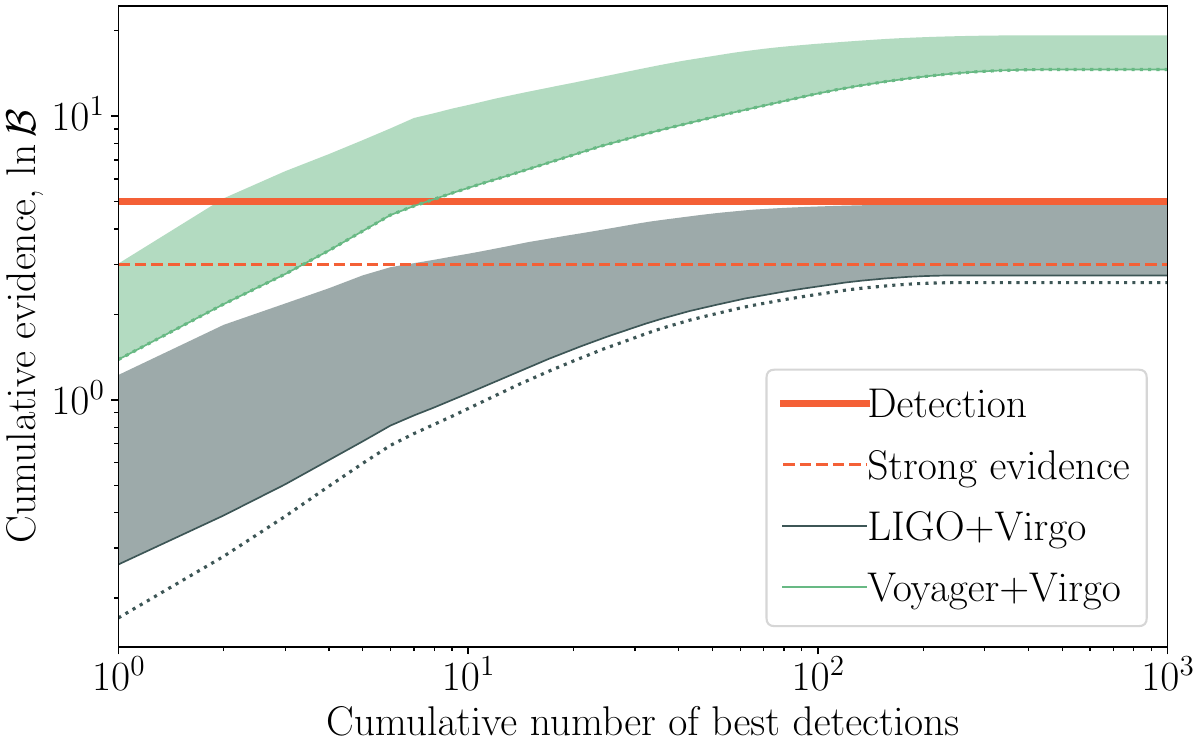}
        \label{fig:combine:lvkevents}
    \end{subfigure}
    \begin{subfigure}[b]{0.49\textwidth}
        \includegraphics[width=\textwidth]{pop_cumulative_etce_time.pdf}
        \label{fig:combine:xgtime}
    \end{subfigure}
    \begin{subfigure}[b]{0.49\textwidth}
        \includegraphics[width=\textwidth]{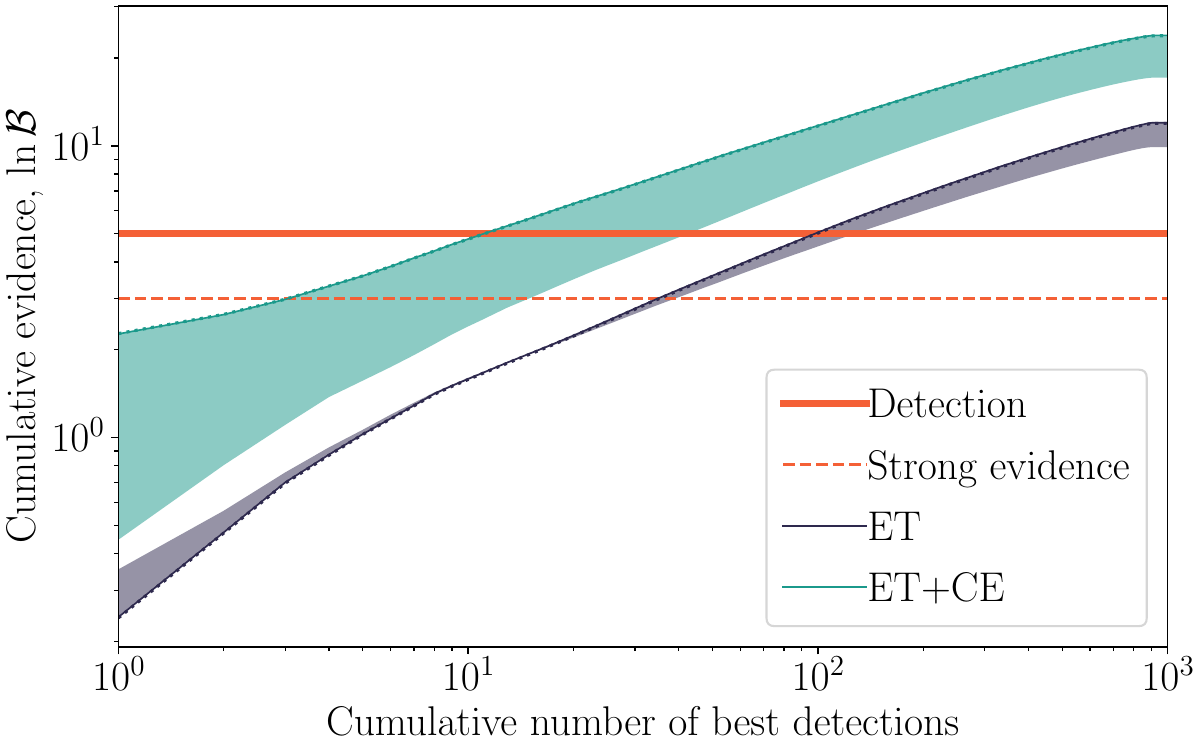}
        \label{fig:combine:xgevents}
    \end{subfigure}
    \vspace{-1.2\baselineskip} 
    \caption{Model selection with ground-based detectors based on $\ln \mathcal{B}$. \textit{Top}: the Poincar{\'e} group versus the original BMS group. \textit{Bottom:} the original BMS versus the extended BMS, the result is only conclusive for next-generation instruments (ET, CE). Solid lines and dotted lines represent simulations of a more complex model and a least complex model of the two, respectively. Evidence for the simulated symmetry group $\ln\mathcal{B}$ is shown as a function of (a) the observation time, the left two panels, and (b) the number of optimal BBH observations, the right two panels.
    } 
    \label{fig:combine_sup}
\end{figure*}

\begin{figure*}[!htb]
    \centering
    \begin{subfigure}[b]{0.32\textwidth}
        \includegraphics[width=\textwidth]{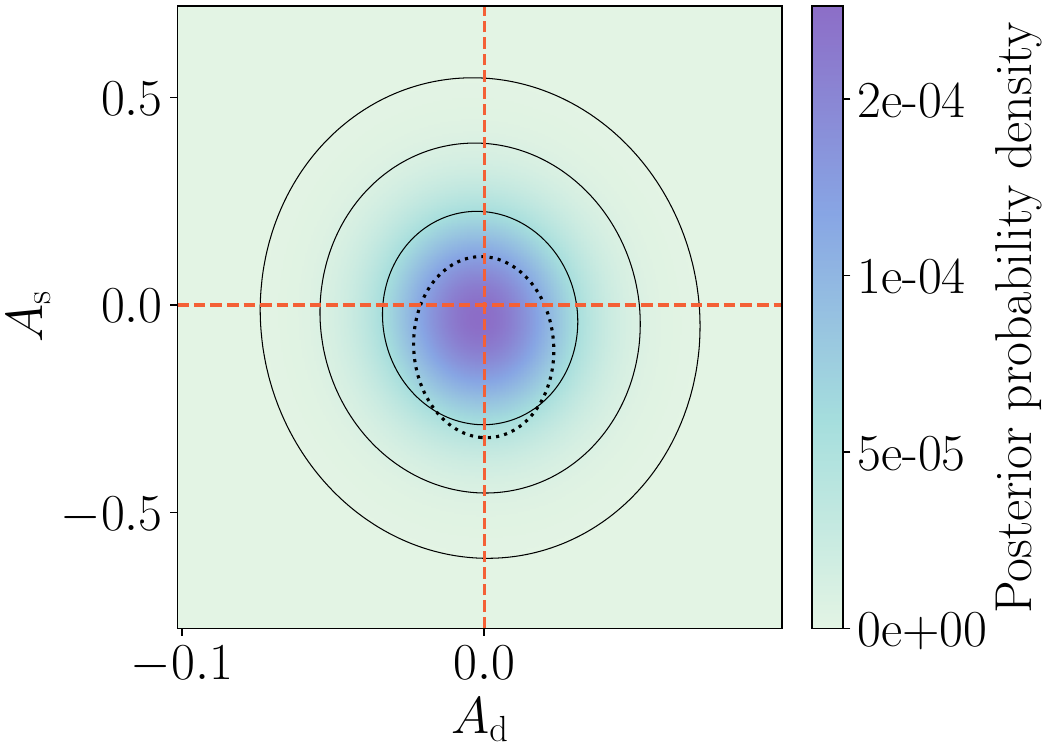}
        \caption{Poincar{\'e} Universe}
        \label{fig:pe_xg:m}
    \end{subfigure}
    \begin{subfigure}[b]{0.32\textwidth}
        \includegraphics[width=\textwidth]{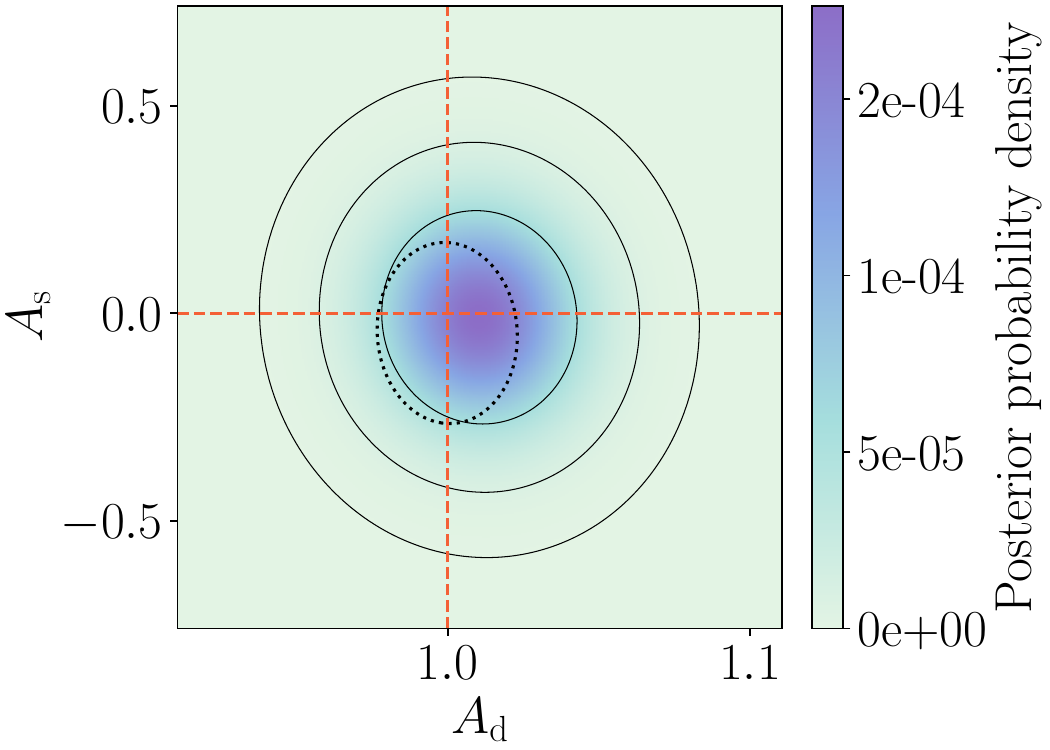}
        \caption{Original BMS Universe}
        \label{fig:pe_xg:nom}
    \end{subfigure}
    \begin{subfigure}[b]{0.32\textwidth}
        \includegraphics[width=\textwidth]{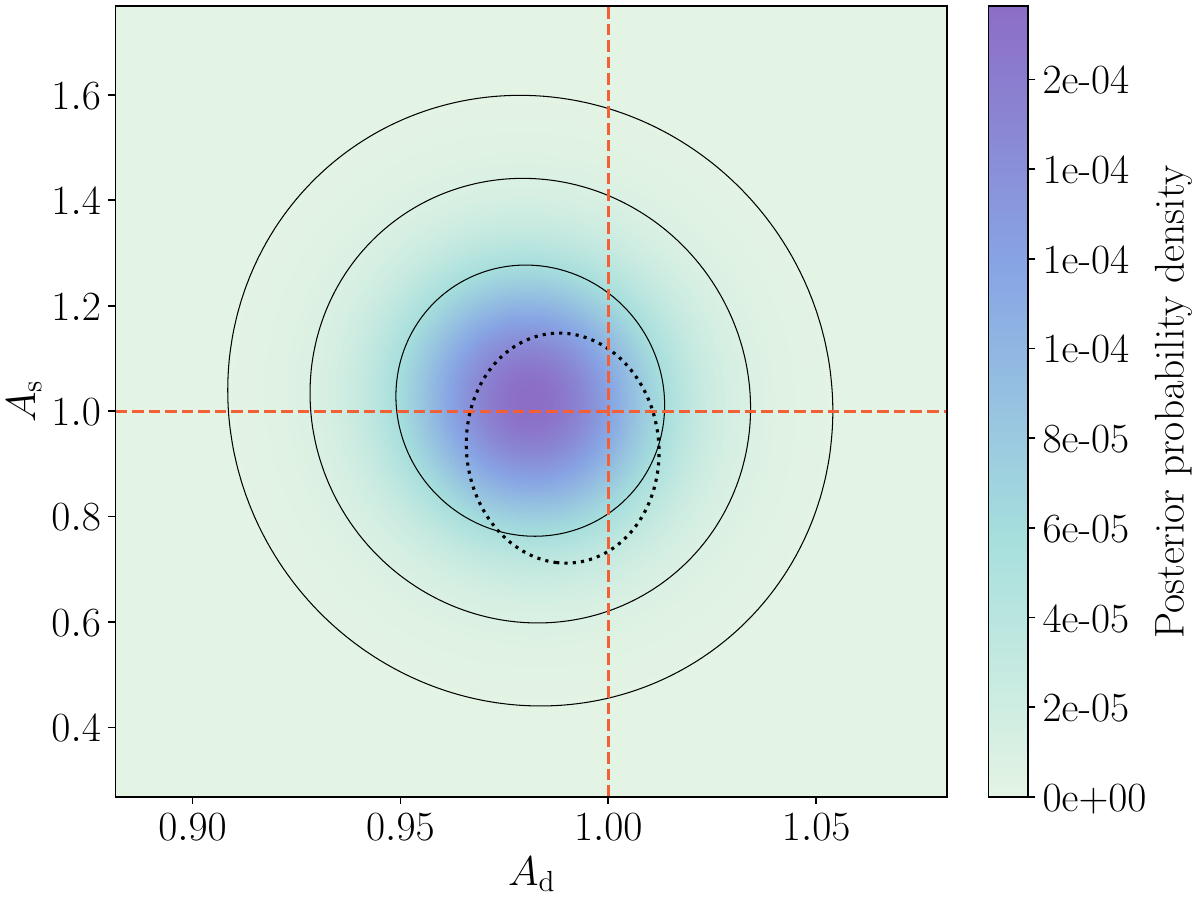}
        \caption{Extended BMS Universe}
        \label{fig:pe_xg:nje}
    \end{subfigure}
    \vspace{-0.5\baselineskip} 
    \caption{Relative memory amplitudes, $A_\text{d,s}$, with $10^3$ optimal BBHs observed by the Einstein Telescope in a year. $A_\text{d,s}$ corresponding to observing scenarios in three panels -- either 0 (memory off) or 1 (memory on) -- are indicated by dashed lines. The color represents posterior probability, $\mathcal{P}(A_\text{d,s}|d)$. Solid lines show $1,2,3$-$\sigma$ credible levels. Dotted lines show $1$-$\sigma$ credible levels when we include all BBH mergers up to a redshift $z \approx 1$ ($10^4$ events).
    } 
    \label{fig:pe_xg_sup}
\end{figure*}

In this Section, we provide an extended set of figures for our Letter. 
For convenience, a visualization of the gravitational wave memory signal is provided in Figure~\ref{fig:mem_wave_sup}. 
Extended results for ground-based interferometers are shown in more detail in Figure~\ref{fig:combine_sup}. 
The results for measuring $A_\text{d,s}$ are extended to simulations of symmetry groups other than the extended BMS in Figure~\ref{fig:pe_xg_sup}.

\section{\label{sec:gr} Subtleties in memory modelling}

In this Letter, we use a hybridized numerical relativity (NR) surrogate approximate of the gravitational waveform $h(t)$~\citep{VarmaField2019}. 
Hybridization combines post-Newtonian and effective-one-body approaches for early inspiral phases of compact binaries and uses NR simulations for merger predictions.
``Surrogate'' indicates interpolation among limited NR simulations for compact binaries.
As a matter of fact, NR simulations play an important role in both the fundamental aspects of gravitational wave memory and the accuracy of our results.
Although we discuss a lack of clarity in the symmetry group of asymptotically flat spacetimes (in GR), NR simulations may show what memory terms arise in GR.
With a few caveats, current NR simulations verify the presence of the displacement and spin memory associated with the extended BMS group. 

The first caveat, both displacement and spin memory are found in NR based on the boundary metric conditions of the generalized BMS. 
The implications of boundary conditions of the extended BMS may be an interesting subject for future work.
Considering also a subtle relation between spin memory and the extended BMS group discussed in the Letter, the non-detection of the symmetries discussed in this work may also be interpreted as a deviation from GR. 
We believe our proposed approach of inferring spacetime symmetries is an important key in verifying the universal ``infrared triangle'' relations, complementary to advances in theory and NR.

The second caveat, historically, memory terms were not recovered immediately due to various issues~\citep{MitmanMoxon2020,MitmanIozzo2021}.
The issues were resolved shortly after we finished our calculations thanks to a new technique to fully evolve Einstein equations to future null infinity~\citep{YooMitman2023}. 
NR simulations underlying our employed waveform approximant lack displacement memory but unexpectedly show about half the expected spin memory strain~\citep{MitmanMoxon2020}.
We calculate memory contributions to $h(t)$ using BMS balance laws~\citep{MitmanIozzo2021}, treating any presumed pre-existing spin memory as part of the memory-free waveform.
Since the artifact is $\sim 10^{-3}$ of the strain, its incorporation into spin memory results in a negligible $\sim 10^{-6}$ strain impact.
This artifact vanishes in the latest surrogate approximant version with integrated memory contributions~\citep{YooMitman2023}.

\section{\label{sec:results} Particularities in the results}

Let us point out a few notable features in the results for ground-based interferometers.
First, we find that the local black hole merger rate is not an important factor at observation timescales of $\sim 100$ days required for the detection of relevant symmetry groups.
In these early days, at some observation intervals, the auxiliary low-rate population even yields higher evidence for a correct symmetry group compared to the primary population based on the best-fit merger rate.
This is especially prominent for LIGO-Virgo and Voyager-Virgo in the context of displacement memory.
This is partly due to the modest contribution of the majority of the signals to the measurement.
As shown in the right upper panel in Figure~\ref{fig:combine_sup}, only 2\% events contribute 50\% of all evidence, whereas the growth of evidence turns over after around $200$ observed black hole mergers.
Whereas for the case of ET probing higher-order BMS symmetries through spin memory, the positive impact of the higher merger rate is more clear.
There, 10\% of the events contribute to 50\% of all evidence, as shown in the right bottom panel in Figure~\ref{fig:combine_sup}.

Second, for ET and CE, the evidence for a correct symmetry group tends to match in value between the cases where the roles of correct and incorrect symmetry groups are switched.
So, the information about the spacetime symmetry group inferred from the presence of the spin memory term is equal to the information inferred from the absence of spin memory in the opposite scenario.
However, there appears to be a discrepancy in the case of displacement memory and observations with LIGO, Virgo, and Voyager.
The dotted line, which represents the case of the simulated original BMS symmetry group, is visually below the solid line, which represents the case of the simulated Poincar{\'e} symmetry group.
The offset can be explained by stronger GW parameter estimation degeneracies for LIGO, Virgo, and Voyager, compared to ET and CE.
These degeneracies make it more difficult to evaluate the Fisher information matrix that governs the posterior, $\mathcal{P}(\theta|d)$. 
In turn, this affects the evidence values, $\mathcal{Z}$, and thus the Bayes factors.

Finally, one may notice a visible leap in evidence between 150 and 175 observation days with ET and CE, compared to ET only.
For ET and CE, the log Bayes factor is increased by $2.3$.
This is due to a significant improvement in parameter estimation thanks to CE, for BBH merger with component masses of $35~M_\odot$ and $27~M_\odot$, with an orbital inclination of $54$ deg. 
The most prominent difference for this event was related to sky position and polarization parameters, which were almost unconstrained for ET.
For this event, it is likely that ET sky position uncertainty is underestimated, as typically Fisher matrix approximation of the posterior is suboptimal for multi-modal probability distributions with nonlinear covariance~\cite[e.g., Figures 6-9 in][]{DupletsaHarms2023}.
Nevertheless, this event is an example of the positive impact of forming a network of next-generation detectors, compared to relying on the observations of individual detectors.

\section{\label{sec:fisher} Reduced-order Bayesian inference}

State-of-the-art Bayesian parameter estimation requires computationally expensive posterior sampling, as typically the functional form of the posterior is unknown.
To be able to study thousands of signals, we Taylor-expand the likelihood at simulated parameter values $\theta_0$ and approximate it by neglecting terms past the second order.
The resulting posteriors will become a multivariate Gaussian distribution centered at true parameter values, with measurement uncertainty given by the inverse of the Fisher information matrix~\citep{DupletsaHarms2023}. 
By knowing the form of the posterior, we also evaluate Bayesian evidence analytically to perform model selection. 
The Fisher matrix approximation is valid and sufficiently accurate for the GW signals we discuss throughout the paper because we verify that the signals are sufficiently loud.
When taking a product of Bayes factors and measuring $A_\text{d,s}$ based on multiple BBHs, we perform marginalization over nuisance parameters.
The only requirement here is that the marginalization is not performed over common parameters between the measurements, such as $A_{\text{d},\text{s}}$ themselves. 

It was also pointed out in~\citep{LaskyThrane2016} that the degeneracy between $\phi$ and $\psi$ in the dominant $(2,2)$ mode of a signal prevents determining the sign of the memory.
This may invalidate the assumption that there is only one maximum-\textit{aposteriori} point in the parameter space so that $\ln\mathcal{B}$ is overestimated.
We use the waveform approximant with high-order modes to break this degenearcy~\citep{VarmaField2019}.
The loudness of the high-order modes is ensured by checking that the optimal SNR of the $m$-odd part of the waveform is $>2$, following~\citep{GrantNichols2023}.
For simplicity, we assign $\ln\mathcal{B}$ to zero for signals that do not meet the criterion.
This results in a slightly more conservative prediction.

\subsection{\label{sec:fisher:likelihood} Likelihood}

Let us provide more details on the use of Fisher matrix approximation of the likelihood in Bayesian inference from our simulations.
We start with Taylor-expanding the log-likelihood, $\ln\mathcal{L}(d|\bm{\theta})$, with respect to the vector of parameters $\bm{\theta}$ at the position of true parameters $\bm{\theta}_0$ that describe the data, $d$:
\begin{equation}
    \ln\mathcal{L}(d|\bm{\theta}) \bigg|_{\bm{\theta} \xrightarrow{} \bm{\theta}_0} = \sum_{n=0}^{\infty} \ln\mathcal{L}^{(n)}(d|\bm{\theta}_0)(\bm{\theta}-\bm{\theta}_0)^{n},
\end{equation}
where $(n)$ denotes the $n$'th derivative of the likelihood with respect to $\bm{\theta}$.
The $n=0$ term corresponds to the maximum (log-)likelihood, $\mathcal{L}_0 = \text{const}(\bm{\theta})$. 
Because the $n=1$ term corresponds to the first derivative of the likelihood at its maximum point, it is equal to zero.
Note that the $n=2$ term contains 
\begin{equation}
    \ln\mathcal{L}^{(2)}(d|\bm{\theta}_0) = \langle \frac{\partial^2}{\partial \theta_i \partial \theta_j}\ln\mathcal{L}(d|\bm{\theta}_0) \rangle \equiv \bm{F},
\end{equation}
which is equal to the Fisher information matrix~\citep{DupletsaHarms2023}, $\bm{F}$, a Hessian matrix where $i$ and $j$ denote indices of the first and the second partial derivatives.
Neglecting terms past the second order, the likelihood is approximated as 
\begin{equation}\label{eq:approx_likelihood}
    \mathcal{L}(d|\bm{\theta}) \bigg|_{\bm{\theta} \xrightarrow{} \bm{\theta}_0} \approx \mathcal{L}_0 \exp \bigg( -\frac{1}{2}(\bm{\theta} - \bm{\theta}_0) \bm{F} (\bm{\theta} - \bm{\theta}_0) \bigg).
\end{equation}
Notice that this approximation represents a multivariate Gaussian function, with a covariance matrix $\bm{C}=\bm{F}^{-1}$ given by the inverse of the Fisher information matrix.
The dependence of the likelihood on $\theta$ is what contributes to the posterior distribution, $\mathcal{P}(\bm{\theta}|d) \propto \mathcal{L}(d|\bm{\theta}) \pi(\bm{\theta})$, representing our simulated measurement uncertainties.
In this work, we assume flat uninformative priors, such that $\pi(\bm{\theta})=1$ and the posterior is determined only by the likelihood.
When true parameters $\bm{\theta}_0$ are not known, \textit{e.g.} the case of real observations, priors should be chosen such that they outline the true distributions of the observed parameters in nature.
However, this is not the case in our study.

\subsection{\label{sec:fisher:selection} Model selection and conditional posteriors}

One may also evaluate the posterior and the evidence conditioned upon fixing one or more of the parameters of the model that were previously free.
Such a posterior will remain a multivariate Gaussian function, a slice of the original multivariate Gaussian posterior.
Let us outline how to find $(\bm{\theta}_0',\bm{C}',\mathcal{L}_0')$, the new mean, the new covariance matrix, and the new maximum likelihood position of such a conditioned posterior, respectively, which fully define it.
Let us split the parameters into the fixed, denoted by an index ``f'', and the remaining ones, denoted by an index ``r'', such that $(\bm{\theta}_\text{f},\bm{\theta}_\text{r})=\bm{\theta}$.
Similarly, we break the true values into $(\bm{\theta}_\text{f,0},\bm{\theta}_\text{r,0})=\bm{\theta}_0$.
When fixing parameters $\bm{\theta}_\text{f}=\bm{a}=\text{const}$, we reduce the dimensionality of the posterior.
To solve for $\bm{\theta}_0'$ and $\bm{C}'$, we break the covariance matrix into so-called Schur complements,
\begin{equation}
\bm{C} =
    \begin{pmatrix}
    \bm{C}_\text{rr} & \bm{C}_\text{rf} \\
    \bm{C}_\text{fr} & \bm{C}_\text{ff}
    \end{pmatrix},
\end{equation}
where $\bm{C}_\text{ff}$ are the elements where rows and columns correspond to the fixed parameters $\bm{\theta}_\text{f}$, $\bm{C}_\text{rr}$ are the elements where rows and columns correspond to the remaining parameters $\bm{\theta}_\text{r}$, $\bm{C}_\text{rf}$ is where rows correspond to the remaining parameters and columns correspond to the fixed parameters, and $\bm{C}_\text{fr}$ is the opposite to the previous one.
Then, it follows that the mean of a conditioned Gaussian function is
\begin{equation}
    \bm{\theta}_0' = \bm{\theta}_\text{r,0} + \bm{C}_\text{rf} \bm{C}_\text{ff}^{-1} (\bm{a} - \bm{\theta}_\text{f,0}).
\end{equation}
The second term in the equation above manifests, in practice, systematic errors in the remaining free parameters $\bm{\theta}_\text{r}$ due to fixing parameters $\bm{\theta}_\text{f}$ at incorrect values.
Whereas the variance of a conditioned Gaussian is 
\begin{equation}
    \bm{C}' = \bm{C}_\text{rr} - \bm{C}_\text{rf} \bm{C}_\text{ff}^{-1} \bm{C}_\text{fr}.
\end{equation}
Note that the parameters can be fixed at both the correct values, $\bm{a} = \bm{\theta}_\text{f,0}$, and the incorrect values, $\bm{a} \neq \bm{\theta}_\text{f,0}$.
In the former case, $\mathcal{L}_0' = \mathcal{L}_0$.
In the latter case, based on Equation~\ref{eq:approx_likelihood}, the new conditional posterior is positioned at $\bm{\theta}_\text{r,0}$ will have a maximum value
\begin{equation}
    \mathcal{L}_0' = \mathcal{L}_0 \exp \bigg( -\frac{1}{2}(\bm{\theta}'' - \bm{\theta}_0) \bm{F} (\bm{\theta}'' - \bm{\theta}_0) \bigg) < \mathcal{L}_0,
\end{equation}
where $\bm{\theta}''=(\bm{a},\bm{\theta}_\text{r,0})$.

Considering nested models, the approach above allows one to evaluate the Bayes factors for one model over the other.
For all calculations throughout this work, we implicitly evaluate the Fisher matrix for $A_\text{d,s}$ in addition to the standard gravitational wave parameters.
Next, following the previous paragraph, we evaluate the posterior for a set of hypotheses where $\theta_\text{f}=A_\text{d,s}$ are fixed to values $0$ and $1$, depending on the spacetime symmetry in question.
Because our approximation of the posterior given by Equation~\ref{eq:approx_likelihood} represents a function for which the integral is known, we evaluate Bayesian evidence analytically,
\begin{equation}
    \mathcal{Z} \equiv \int \mathcal{L}(d|\bm{\theta}) \pi(\bm{\theta}) d\bm{\theta} = \mathcal{L}_0 \sqrt{(2\pi)^m \det{\bm{C}}},
\end{equation}
where $m$ is a number of parameters in vector $\bm{\theta}$, the dimensionality of a multivariate Gaussian.
Similarly, we evaluate the evidence $\mathcal{Z}'$ for the nested models that depend on $(\bm{\theta}_0',\bm{C}',\mathcal{L}_0')$.
Finally, we evaluate a Bayes factor in favor of model A against model B as $\mathcal{B} \equiv \mathcal{Z}_\text{A}/\mathcal{Z}_\text{B}$.

\subsection{\label{sec:fisher:valid} On the accuracy of the approximation}

A number of caveats apply to the use of Fisher matrix approximation of the likelihood.
In particular, the accuracy may be subject to sufficiently high signal-to-noise ratios (SNRs) for the signals of interest, the inclusion of priors, as well as the numerical accuracy of the matrix inversion~\citep{Vallisneri2008}.
Except for the LIGO-Virgo simulations, for which we point out the effect of a weak-signal case in Section~\ref{sec:results}, we work in the very high SNR regime as we focus on a sub-population of the loudest BBH mergers.
The effect manifests as a $\approx 10\%$ discrepancy between $\ln\mathcal{B}^\text{Poincar{\'e}}_\text{BMS}|_\text{Poincar{\'e}}$ and $\ln\mathcal{B}^\text{BMS}_\text{Poincar{\'e}}|_\text{BMS}$ in Figure~\ref{fig:combine_sup}.
Note, the effect is still smaller than the uncertainties associated with the (1) BBH merger rate, (2) cosmic variance.
At the ballpark, our results are consistent with full Bayesian inference by~\citep{HubnerTalbot2020}.
They find that LIGO-Virgo requires around $1500$ BBH mergers to detect the displacement memory at $\ln\mathcal{B}=5$.
Approximately the same amount of events is to be detected by LIGO-Virgo at design sensitivity in a year, which is equal to our detection time scale.
Note that it was also found in~\citep{HubnerTalbot2020} that the result is largely subject to cosmic variance, which we did not simulate for ground-based detectors.
Nevertheless, the effect of cosmic variance is apparent in Figure~\ref{fig:combine_sup} when at some point during the observation the cumulative evidence in favor of a correct symmetry group given the pessimistic black hole merger rate exceeds that obtained with the best-fit merger rate.
There are also more optimistic and more pessimistic prospects in the literature.
Shortly after the first gravitational wave detection, Lasky \textit{et al}.~\citep{LaskyThrane2016} suggested that LIGO may be able to detect the displacement memory in 90 days at SNR of 5, assuming all events are like GW150914.
Although the assumption is now considered too optimistic, our detection timescale for the displacement memory with LIGO-Virgo is larger only by a factor of four.
Regarding the prospects of detection of both the displacement and the spin memory, our detection timescales are shorter compared to~\citep{GrantNichols2023} by around a factor of 3.
This may be due to differences in the methodology and partly to due to consideration of different detector networks.
In any case, we argue that the level of precision of the Fisher matrix approximation is inferior to other uncertainties for the purpose of our study.

\section{\label{sec:optimistic} Optimistic and pessimistic assumptions}

The predictive accuracy of our results is subject to a number of both conservative and optimistic assumptions.
The conservative assumptions are the following.
First, we neglect the fact that some black holes have non-negligible spins, although non-zero spins increase memory amplitude.
Secondly, we only used the subset of loudest events, where in practice it is easier to control the noise and systematic errors.
It is not so straightforward in the analysis of significant yet relatively memory-quiet events (e.g., face-on), although including these is likely to improve the measurement.
The use of quieter signals would require a full Bayesian inference. 
While for LIGO-Virgo considering quieter signals does not add to our knowledge because of a lack of high-order mode content, for ET and CE there is more information about memory that can be extracted from the data (see also Figure 4 where we show the difference between $10^3$ BBHs and $10^4$ BBHs). 
Because BBHs are selected based on the optimal SNR for spin memory, they are further only sub-optimal for resolving the displacement memory term.
Furthermore, we do not discuss binaries that contain neutron stars as they are typically quieter.
Among the optimistic assumptions, we note that it is not guaranteed that the instruments will reach design sensitivity.
Additionally, data quality issues and non-Gaussian noise in LIGO and Virgo are known to limit the ability to interpret the astrophysical origin of signals~\citep{PayneHourihane2022}.
In principle, it is also possible that our method of measuring memory strain amplitude may be sub-optimal for the identification of certain cases of a misspecified GW memory model.

\vspace{1\baselineskip} 
\section{Data and code availability}

The code to reproduce the results of our analysis and the simulated data is available at \href{https://github.com/bvgoncharov/gwmem_2022}{github.com/bvgoncharov/gwmem\_2022}.
The response of detectors to gravitational wave signals and the calculations of the Fisher matrix and the error matrix are performed with \href{https://github.com/bvgoncharov/GWFish/tree/development_bg}{this branch} of \textsc{gwfish}~\citep{DupletsaHarms2023}.
To calculate the gravitational wave strain for a given event, we employ the numerical relativity waveform approximant \textsc{\MakeLowercase{NRHybSur3dq8}}~\citep{VarmaField2019} available through \textsc{gwsurrogate}.

\end{document}